\documentclass[a4paper,11pt]{article}
\usepackage{jinstpub} % for details on the use of the package, please see the JINST-author-manual
\usepackage{subcaption}
\usepackage{multirow}

\title{\boldmath Spatial resolution improvement of PICOSEC Micromegas precise timing detectors}

\author[a,1]{F.M. Brunbauer \note{Corresponding author}}
\author[b]{R. Aleksan}
\author[c]{Y. Angelis}
\author[b]{S. Aune}
\author[e]{J. Bortfeldt}
\author[f,g]{M. Brunoldi}
\author[h]{J. Datta}
\author[b]{D. Desforge}
\author[i]{G. Fanourakis}
\author[f,g,2]{D. Fiorina\note{Now at Gran Sasso Science Institute, Viale F. Crispi, 7 67100 L'Aquila, Italy}}
\author[a]{K. J. Floethner}
\author[k]{M. Gallinaro}
\author[l]{F. Garcia}
\author[b]{I. Giomataris}
\author[m]{K. Gnanvo}
\author[o]{Q. Huang}
\author[b,3]{F.J. Iguaz\note{Now at SOLEIL Synchrotron, L'Orme des Merisiers, 91190 Saint Aubin, France}}
\author[a]{D. Janssens}
\author[b]{A. Kallitsopoulou}
\author[c]{I. Karakoulias}
\author[n]{M. Kovacic}
\author[b]{P. Legou}
\author[a]{M. Lisowska}
\author[o]{J. Liu}
\author[p]{M. Lupberger}
\author[b,c,4]{I. Maniatis\note{Now at Department of Particle Physics and Astronomy, Weizmann Institute of Science, Rehovot, Israel.}}
\author[n]{M. Micetic}
\author[a,p]{H. Muller}
\author[a]{E. Oliveri}
\author[b]{T. Papaevangelou}
\author[r]{M. Pomorski}
\author[a]{L. Ropelewski}
\author[n]{K. Salamon}
\author[c,d]{D. Sampsonidis}
\author[a]{L. Scharenberg}
\author[a]{T. Schneider}
\author[r]{E. Scorsone}
\author[b,5]{L. Sohl\note{Now at TUV NORD EnSys GmbH \& Co. KG.}}
\author[h]{N. Shankman}
\author[a]{M. van Stenis}
\author[s]{Y. Tsipolitis}
\author[c,d]{S.E. Tzamarias}
\author[t]{A. Utrobicic}
\author[f,g]{I. Vai}
\author[a]{R. Veenhof}
\author[f,g]{P. Vitulo}
\author[o]{X. Wang}
\author[u]{S. White}
\author[o]{Z. Zhang}
\author[o]{Y. Zhou}

\affiliation[a]{CERN, 1211 Geneva 23, Switzerland}

\affiliation[b]{IRFU, CEA-Université Paris-Saclay, F-91191 Gif-sur-Yvette, France}
\affiliation[c]{Department of Physics, Aristotle University of Thessaloniki, GR-54124 Thessaloniki, Greece}
\affiliation[d]{CIRI-AUTH, GR-57001 Thessaloniki, Greece}
\affiliation[e]{Department for Medical Physics, Ludwig Maximilian University of Munich, 85748 Garching, Germany}

\affiliation[f]{Dipartimento di Fisica, Università di Pavia, 27100 Pavia, Italy}
\affiliation[g]{INFN Sezione di Pavia, 27100 Pavia, Italy}
\affiliation[h]{Department of Physics and Astronomy, Stony Brook University, NY 11794-3800, USA}
\affiliation[i]{Institute of Nuclear and Particle Physics, NCSR Demokritos, GR-15341 Agia Paraskeui, Attiki, Greece}
\affiliation[k]{Laboratório de Instrumentação e Física Experimental de Partículas (LIP), Lisbon, Portugal}
\affiliation[l]{Helsinki Institute of Physics, University of Helsinki, FI-00014 Helsinki, Finland}
\affiliation[m]{Jefferson Lab, Newport News, VA 23606, USA}
\affiliation[n]{Faculty of Electrical Engineering and Computing, University of Zagreb, 10000 Zagreb, Croatia}
\affiliation[o]{State Key Laboratory of Particle Detection and Electronics, University of Science and Technology of China, 230026 Hefei, China}
\affiliation[p]{Physikalisches Institut, University of Bonn, 53115 Bonn, Germany}
\affiliation[r]{CEA-List, Diamond Sensors Laboratory, CEA-Saclay, F-91191 Gif-sur-Yvette, France}
\affiliation[s]{National Technical University of Athens, Athens, Greece}
\affiliation[t]{Ruđer Bošković Institute, 10000 Zagreb, Croatia}
\affiliation[u]{University of Virginia, Virginia, USA}

\emailAdd{florian.brunbauer@cern.ch}

\abstract{
The combination of a Cherenkov radiator with a semi transparent photocathode and a Micromegas based amplification stage allows PICOSEC Micromegas detectors to achieve a time resolution of better than 15\,ps. While tileable prototypes with 10x10 channels feature $1 x 1\,cm^{2}$ readout pads, finer readout granularity can be used to improve the spatial resolution. We report on the study of high readout granularity PICOSEC Micromegas prototypes which achieve around 0.5\,mm spatial resolution with 3.5\,mm large pads. No significant improvement was found when readout pad size was further reduced to 2.2\,mm. The timing resolution of the leading pad was found to be slightly degraded but remained better than 20\,ps for a medium granularity prototype. The achieved spatial resolution can enable PICOSEC Micromegas to be used as precise timing and moderate resolution tracking detector simultaneously.
}

\keywords{Gaseous detectors, MPGD, precise timing, Micromegas, spatial resolution, timing resolution}

\begin{document}

\maketitle

\flushbottom

\section{Introduction}
\label{sec:intro}

\begin{figure}[h!]
\centering
\makebox[\linewidth]{\includegraphics[width=0.7\textwidth]{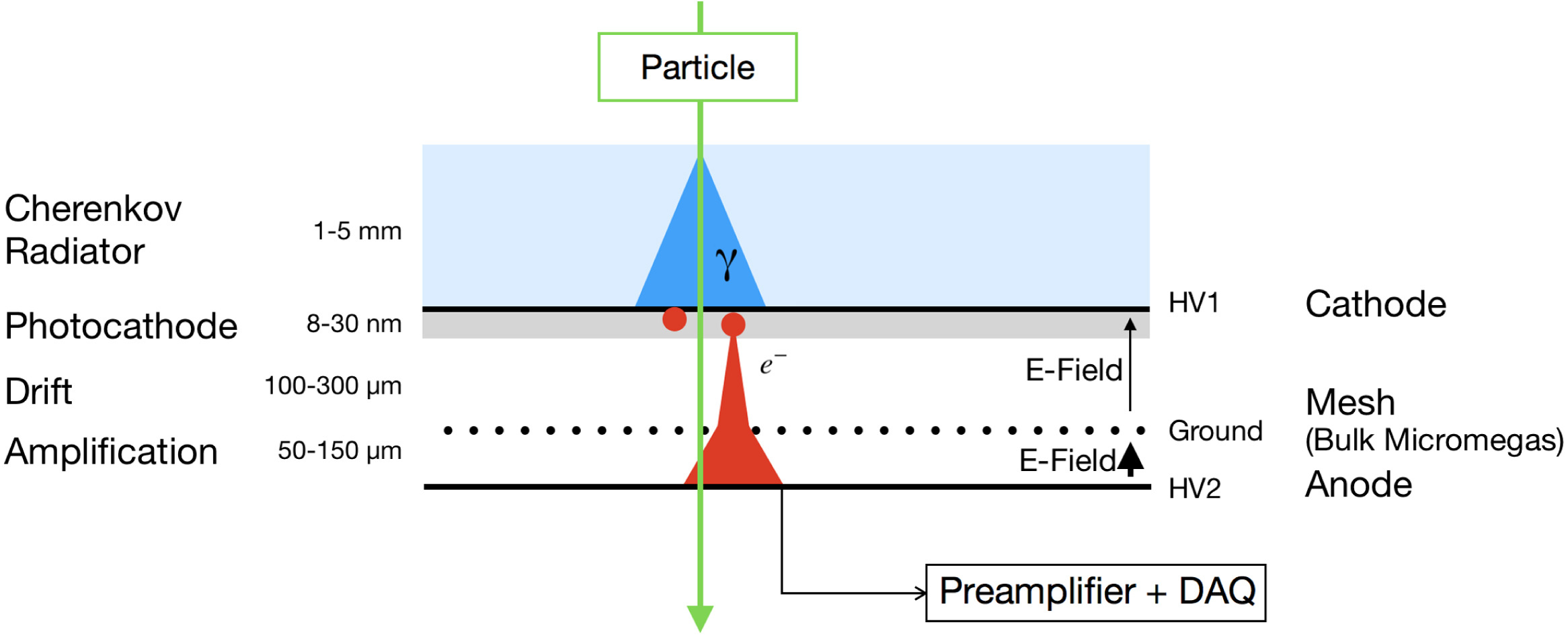}}
\caption{Picosec Micromegas schematic: Cherenkov radiator with semi-transparent photocathode and two-stage Micromegas amplification structure. Taken from \cite{BORTFELDT2018317}.}
\label{picosecSchematic}
\end{figure}

Precise timing detectors are important for future high energy physics experiments for multiple reasons. The can be used for pileup mitigation in high multiplicity environments, provide accurate TOF information for accurate PID and can be used for 4D tracking. These applications would profit from timing resolution on the order of tens of picoseconds. While gaseous detectors are in wide use as large active area detectors and provide e.g. tracking information in muon systems, they are typically not associated to high timing resolution. Typically, the uncertainty in the location of primary ionisation in the active region results in a timing jitter on the order of nanoseconds which cannot be easily overcome with a millimetre-scale conversion region thickness.
The PICOSEC Micromegas concept combined a Cherenkov radiator with a semi-transparent photocathode and an amplification stage based on MicroPattern Gaseous Detector (MPGD) technologies as shown in figure \ref{picosecSchematic} to achieve better than 20ps timing resolution \cite{BORTFELDT2018317}. 

These detectors make use of the localisation of the primary charge production in space and time at the photocathode as well as thin pre-amplification and amplification gaps with high electric fields. These are used to start avalanche multiplication as soon as possible to minimise the influence of the primary electrons’ movement and profit from statistical averaging over a larger number of charge carriers.
Previous studies of PICOSEC detectors with Micromegas and µRWELL amplification charges have achieved time resolution values better than 15ps on small, single readout pad prototypes \cite{utrobicic2024singlechannelpicosecmicromegas} and better than 20ps on multi-channel prototypes with 100 channels \cite{Utrobicic_2023}. The tillable multichannel prototype modules feature 1x1 cm large readout pads which are individually readout. One possible electronic readout chain with custom developed RF amplifiers and the SAMPIC waveform TDC as digitiser has been shown to manage the readout of full detector modules without significant degradation in timing resolution.
In the following, we describe the achievable spatial resolution of PICOSEC Micromegas detectors and strategies to improve it which include charge sharing and finer readout granularity.

\section{Experimental methods}
\label{sec:setup}

\begin{figure}[h!]
\centering
\makebox[\linewidth]{\includegraphics[width=0.7\textwidth]{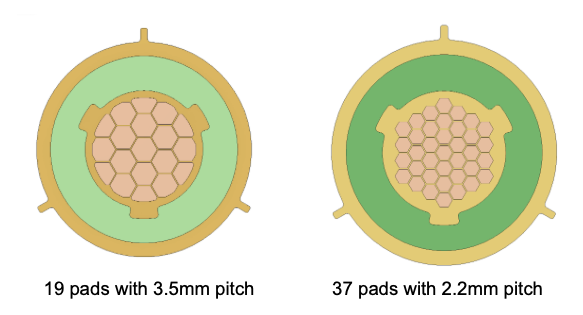}}
\caption{Comparison of Picosec readout structures: medium granularity (left) and high granularity (right). The active area is 15\,mm diameter in both versions.}
\label{readoutGranularityComparison}
\end{figure}

The spatial resolution of three different geometries of PICOSEC Micromegas precise timing detectors was studied: a multipad detector with square 1\,cm x 1\,cm pads, a smaller so-called “medium granularity” prototype with hexagonal pads at a pitch of 3.5\,mm and a “high granularity” prototype with hexagonal pads at a pitch of 2.2\,mm, all produced at the CERN Micro-Pattern Technologies workshop (MPT). A comparison of the two readout granularities is shown in figure \ref{readoutGranularityComparison}.

 The Cherenkov radiator was a 3\,mm thick MgF2 crystal with a semitransparent photocathode coated on it by thermal evaporation. The photocathode coating consisted of a 3nm thick Ti layer for electrical contact and an 18\,nm thick CsI photo conversion layer. This photocathode has previously been shown to provide >10 photoelectrons per Minimum Ionising Particle (MIP). With a Cherenkov angle close to 45$\deg$, a light cone diameter of about 6\,mm diameter is expected for MIPs crossing the 3\,mm thick radiator crystal.
The amplification stage for the different prototypes was a resistive bulk Micromegas with a resistive layer based on Diamond-Like Carbon (DLC) with a nominal sheet resistivity of 20\,M$\Omega$/sq. This high resistivity was chosen to offer protection against discharges while minimally influencing the signal shape and thus timing resolution. The preampfification stage was 180\,µm for the multi-pad detector prototype and 120\,µm for the medium and high granularity prototypes. The mesh was fixed by coverlay pillars defining an amplification gap thickness of around 128\,µm. Detectors were operated in a Ne-CF4-Ethane (80/10/10\,\%) gas mixture at ambient pressure in flushing mode with about 2\,l/h flow. With the cathode at negative high-voltage and the mesh grounded, the resistive layer used as anode was powered with positive high-voltage. Readout pads were separated from the anode by a 50\,µm thick insulating polyimide layer. 
Prototypes were studied in muon beams at the CERN SPS H4 beamline with a momentum of 150\,GeV/c. Timing performance was characterised with a telescope based on three triple-Gaseous Electron Multiplier (GEM) tracking detectors and a MCP-PMT as timing reference. The GEM telescope achieved a tracking resolution of better than 100\,µm. The MCP-PMT provides better than 5\,ps timing resolution in the central region of its active area and is thus a minimal contribution to the overall timing resolution measured. The contribution of the MCP-PMT reference detector was not subtracted from the reported time resolutions. The MCP-PMT signal was split to be used as trigger and timing reference signal simultaneously. 
Custom preamplifiers based on a RF pulse amplifier design from \cite{Hoarau_2021} were connected to the readout pads to preamplify signals before digitising them either at 10\,GS/s with oscilloscopes or at 8.4\,GS/s with the SAMPIC Waveform Time-To-Digital (WTDC) digitiser. A SAMPIC system capable of reading out up to 128 channels was used with each channel triggering independently. The trigger threshold was set to 20\,mV below the baseline, which was set at 1 V to obtain the maximum dynamic range for the negative signal from the readout pads.
While the trigger and timing reference MCP-PMT was kept in a fixed, aligned position for the medium and high granularity detector, it was scanned across and area of multiple pads for the multipad prototype.

\section{Spatial and timing resolution measurements}
\label{sec:spatialRes}

The measurement of spatial and timing resolution with the different readout granularity prototypes are presented in the following subsections. Specific details of measurement conditions and results for each prototype are provided.

\subsection{Multi-pad module prototypes}
\label{sec:spatialResMultipad}
A detector with 100 square pads with a pitch of 10 mm was characterised. A region of 4 pads in a 2x2 matrix was used for spatial resolution measurements. For a Cherenkov cone diameter of 6 mm, 1-4 pads are expected to see a signal for each MIP.

A spatial resolution, defined as the RMS of the distribution of residuals between the position reconstructed by a center-of-gravity method and the known particle track location from the tracking telescope, was measured to be 2.9\,mm (X) and 2.5\,mm (Y) \cite{guerraSummerStudentReport}. The timing resolution between a single pad with the highest amplitude of the detector under test and the MCP-PMT as timing reference was measured to be 45 ps \cite{guerraSummerStudentReport}.

\subsection{Medium readout granularity}
\label{sec:spatialMediumRes}
\begin{figure}[h!]
\centering
\makebox[\linewidth]{\includegraphics[width=0.4\textwidth]{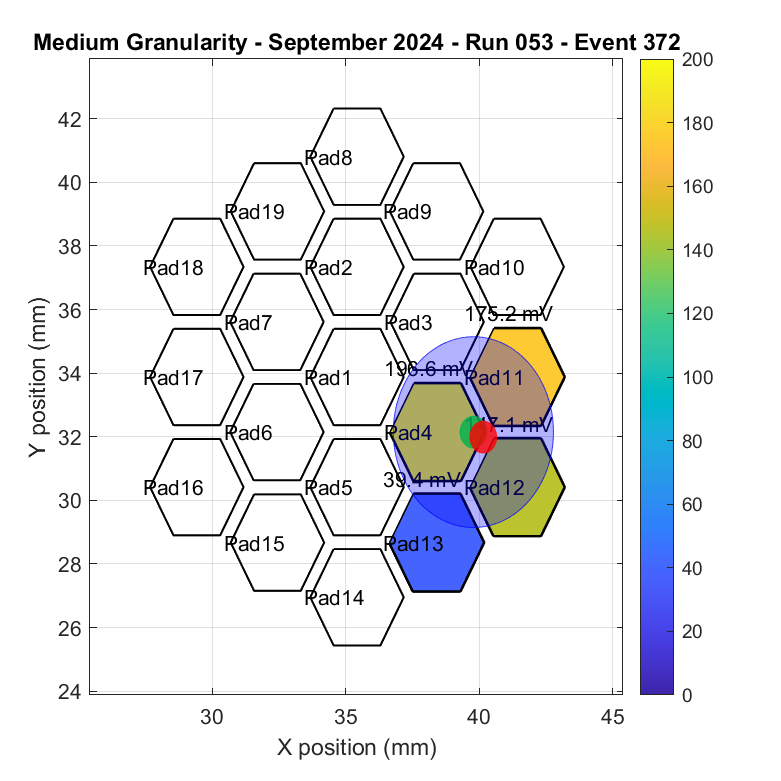}}
\caption{Exemplary event recorded with the medium granularity detector. Four pads are hit, the green point displays the reference tracker information, the semi-transparent blue circle shows the extent of the Cherenkov cone and the red dot shows the reconstructed position from a center-of-gravity algorithm.}
\label{medGranularityEvent}
\end{figure}

A detector with 19 hexagonal pads with a pitch of 3.5 mm was characterised. The total size of the active area of the detector was 15\,mm diameter. For a Cherenkov cone diameter of 6 mm, around 4-7 pads are expected to see a signal for each MIP. 
An exemplary event of a MIP event recorded by the medium granularity detector is shown in figure \ref{medGranularityEvent} with the color scale representing the signal amplitude registered at each pad. As shown, most pads overlapping with the Cherenkov code were hit and recorded signals. 

\begin{figure*}[t!]
    \centering
    \begin{subfigure}[t]{0.5\textwidth}
        \centering
        \includegraphics[height=2in]{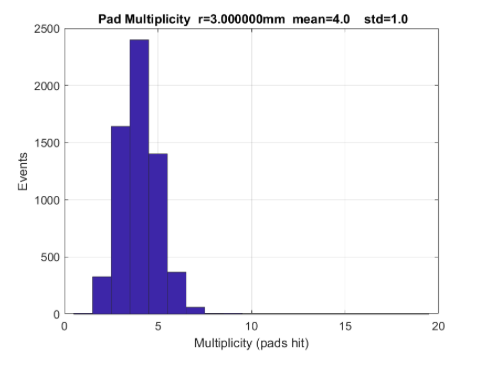}
        \caption{Number of hit pads for medium granularity readout.}
        \label{padMultiplicityMed}
    \end{subfigure}%
    ~ 
    \begin{subfigure}[t]{0.5\textwidth}
        \centering
        \includegraphics[height=2in]{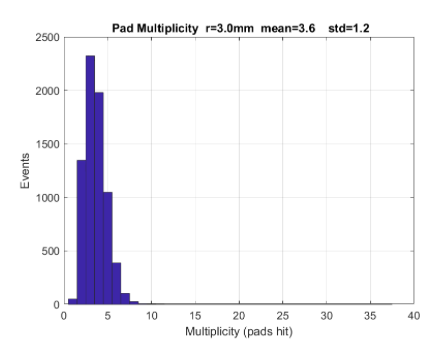}
        \caption{Number of hit pads for high granularity readout.}
        \label{padMultiplicityHigh}
    \end{subfigure}
    \caption{Distribution of the number of hit pads per event for the two readout granularities.}
    \label{fig:horVerProfiles}

\end{figure*}

The distribution of the number of hit pads per event is shown in figure \ref{padMultiplicityMed} which shows an average value of 4.0.

\begin{figure}[h!]
\centering
\makebox[\linewidth]{\includegraphics[width=0.6\textwidth]{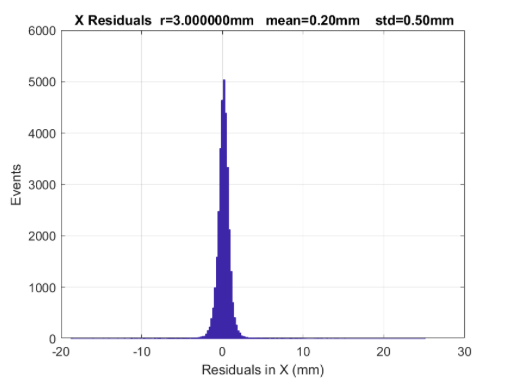}}
\caption{Residual distribution in X direction between measured and reference hit position of the medium granularity readout.}
\label{xResidualsMed}
\end{figure}

\begin{figure*}[t!]
    \centering
    \begin{subfigure}[t]{0.4\textwidth}
        \centering
        \includegraphics[height=2in]{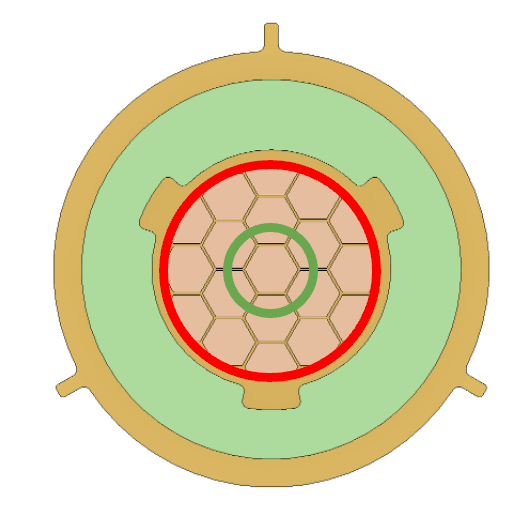}
        \caption{Medium granularity readout selection.}
        \label{centerCircleMed}
    \end{subfigure}%
    ~ 
    \begin{subfigure}[t]{0.4\textwidth}
        \centering
        \includegraphics[height=2in]{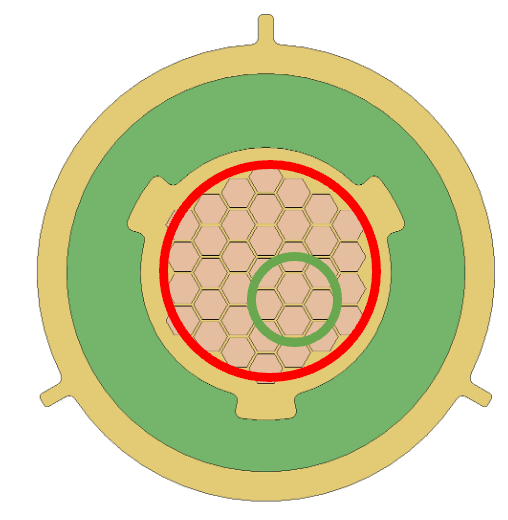}
        \caption{High granularity readout selection.}
        \label{centerCircleHigh}
    \end{subfigure}
    \caption{Overlay of a 6\,mm diamter circle in green displaying the region of the active area which was evaluated for timing and spatial resolution. The red circle shows the full active area.}
    \label{fig:horVerProfiles}

\end{figure*}

Due to the limited active area of the detector, signals outside of a 9\,mm diameter circle were not fully contained and the spatial resolution measurement was thus focused on a central region of the detector with a diameter of 6\,mm as shown by the red circle in figure \ref{centerCircleMed}. A spatial resolution, defined as the RMS of the distribution of residuals between the reconstructed position and the known particle track location from the tracking telescope, was measured to be 0.50\,mm as shown in figure \ref{xResidualsMed}. The resolution in X and Y directions is comparable.

The timing resolution for tracks crossing the central pad and considering only this pad with the highest amplitude was measured to be 16.9±0.1\,ps.

\subsection{High readout granularity}
\label{sec:spatialResHighGranularity}
A detector with 37 hexagonal pads with a pitch of 2.2\,mm was characterised. The total size of the active area of the detector was 15\,mm diameter. For a Cherenkov cone diameter of 6\,mm, around 7-12 pads are expected to see a signal for each MIP.
%An exemplary event of a MIP event recorded by the high granularity detector is shown in figure XXX with the color scale representing the signal amplitude registered at each pad. 

However, in many events multiple pads overlapping with the Cherenkov code were not recording any signal as signal amplitudes on those channels were probably below the self-trigger threshold of the SAMPIC WTDC set at 20\,mV below baseline. The distribution of the number of hit pads per event is shown in figure \ref{padMultiplicityHigh} gives and average value of 3.6 pads hit.

A spatial resolution, defined as the RMS of the distribution of residuals between the reconstructed position and the known particle track location from the tracking telescope, was measured to be 0.65\,mm. The timing resolution for tracks crossing the central pad and considering only this pad with the highest amplitude was measured to be 28.3±0.3\,ps.
Due to the limited active area of the detector, signals outside of a 9\,mm diameter circle were not fully contained and the spatial resolution measurement was thus focused on region of the detector with a diameter of 6\,mm as shown by the red circle in figure \ref{centerCircleHigh}. In contrast to the medium granularity detector, this region was not centered but shifted to the bottom-right by about 2\,mm in each direction to avoid a non-responsive pad.

\section{Discussion}
\label{sec:discussion}

While the average number of hit pads for the medium granularity readout geometry is 4.0 and lies within the expected range, the average number of hits pads for the high readout granularity is even lower at 3.6 despite the finer readout pad pitch. At the same time, compared to the mean amplitudes of the multipad prototype and the medium granularity detector, which is well above 100\,mV, the mean amplitude of 62.3\,mV of signals from the high granularity variants is significantly lower.  With the self-triggering threshold of the SAMPIC WTDC set at 20\,mV, a significant fraction of signals below this threshold may not be recorded. Due to the noise level of the amplifier and inherent noise of the SAMPIC WTDC, it was not possible to set a significantly lower threshold. Thus, the lower number of average pads hit is lower for the high granularity prototype as some pads are below the self-triggering threshold. An alternative readout mode could be a centrally triggered mode in which a single pad above a certain threshold triggers the readout of all readout channels. This readout mode will be evaluated in future tests.
An alternative to modifying the readout mode, would be the operation of the detector at higher gains to ensure that even signals originating from a single photoelectron are large enough to trigger readout of a pad. The preampfification and amplification electric fields at which the detectors were operated were the highest stable values at each configuration thus no margin is available in the same gas mixture. However, studies have shown that a gas mixture of Ne/Isobutane (85/15\,\%) can achieve significantly higher gain and may thus be interesting to achieve higher signal amplitudes with the high granularity geometry.
In addition to providing the hit location of MIPs with better than millimetre accuracy, the demonstrated spatial resolution can also be used to apply advanced corrections for possible non-uniform detector response to further improve timing resolution. No such corrections were applied in the presented results but previous studies which showed a limitation of timing resolution due to non-uniform preamplification gap thickness could profit from position-dependent corrections of the signal-arrival-time (SAT).
As an alternative to the signal sharing due to the extent of the Cherenkov light cone over multiple readout pads, resistive or capacitive signals sharing may be explored. In the case of resistive signal sharing, significantly lower anode resistivity values would be employed to spread signals across multiple readout pads. In addition to decreasing the protective effect of resistive discharge protection, lower resistivity values would also result in slower signal characteristics due to a modification of the delayed signal component and may impact the achievable timing resolution negatively. Capacitive charge sharing is being explored for PICOSEC detectors including Micromegas and µRWELL detector variants.
In addition to the three readout geometries presented in the previous sections, large hexagonal readout pads with an outer diameter of 10 mm were previously evaluated. With a 10\,M$\Omega$/sq resistivity detector with, a spatial resolution of 1.190±0.003\,mm in both X and Y directions was achieved \cite{kallitsopoulou2025performanceoptimizationcharacterization7pad, kallitsopoulou:tel-05267379}. 

\section{Conclusions}
\label{sec:conclusions}

Three different readout pad geometries and granularities were compared for the achievable spatial and timing resolution. A spatial resolution down to 0.5\,mm was achieved with a medium readout granularity with a pitch of 3.5\,mm, while a further increase of the readout granularity did not achieve better results likely due to a limitation of noise and the self-triggering readout mode.
The timing resolution in the single leading amplitude readout pad was slightly lower than the value achieved with a single pad detector which may be explained by a part of the signal being recorded by neighbouring pads. Timing resolution measurements which take advantage of time of signal information in all active pads may further improve the measured resolution. 
The presented studies show that good spatial resolution can be achieved with PICOSEC Micromegas detectors while preserving high timing resolution below 20 ps. Further optimisations in the readout chain and possible operation and higher gains in a different gas mixture will be explored.

%\acknowledgments

% Bibliography
\bibliographystyle{JHEP}
 \bibliography{biblio.bib}

@phdthesis{kallitsopoulou:tel-05267379,
  TITLE = {{Development of a PICOSEC-Micromegas Detector for ENUBET}},
  AUTHOR = {Kallitsopoulou, Alexandra},
  URL = {https://theses.hal.science/tel-05267379},
  NUMBER = {2025UPASP051},
  SCHOOL = {{Universit{\'e} Paris-Saclay}},
  YEAR = {2025},
  MONTH = Jul,
  TYPE = {Theses},
  HAL_ID = {tel-05267379},
  HAL_VERSION = {v1},
}

@misc{kallitsopoulou2025performanceoptimizationcharacterization7pad,
      title={Performance Optimization and Characterization of 7-pad Resistive PICOSEC Micromegas Detectors}, 
      author={A. Kallitsopoulou and R. Aleksan and S. Aune and J. Bortfeldt and F. Brunbauer and M. Brunoldi and J. Datta and D. Desforge and G. Fanourakis and D. Fiorina and K. J. Floethner and M. Gallinaro and F. Garcia and I. Giomataris and K. Gnanvo and F. J. Iguaz and D. Janssens and F. Jeanneau and M. Kovacic and B. Kross and P. Legou and M. Lisowska and J. Liu and M. Lupberger and I. Maniatis and J. McKisson and Y. Meng and H. Muller and E. Oliveri and G. Orlandini and A. Pandey and T. Papaevangelou and M. Pomorski and E. F. Ribas and L. Ropelewski and D. Sampsonidis and L. Scharenberg and T. Schneider and E. Scorsone and L. Sohl and M. van Stenis and Y. Tsipolitis and S. E. Tzamarias and A. Utrobicic and I. Vai and R. Veenhof and P. Vitulo and X. Wang and S. White and W. Xi and Z. Zhang and Y. Zhou},
      year={2025},
      eprint={2512.04842},
      archivePrefix={arXiv},
      primaryClass={physics.ins-det},
      url={https://arxiv.org/abs/2512.04842}, 
}

@techreport{guerraSummerStudentReport,
    author = {Guerra, Francesco and Lisowska, Marta and Oliveri, Eraldo and Brunbauer, Florian},
    title = {PICOSEC Micromegas gaseous detectors for precise timing and developments towards applicable detector},
    institution = {CERN Summer Student Report} ,
    year = {2025}
}

@article{Hoarau_2021,
doi = {10.1088/1748-0221/16/04/T04005},
url = {https://doi.org/10.1088/1748-0221/16/04/T04005},
year = {2021},
month = {apr},
publisher = {IOP Publishing},
volume = {16},
number = {04},
pages = {T04005},
author = {Hoarau, C. and Bosson, G. and Bouly, J.-L. and Curtoni, S. and Dauvergne, D. and Everaere, P. and Gallin-Martel, M.-L. and Marcatili, S. and Muraz, J.-F. and Portier, A. and Rosuel, N.},
title = {RF pulse amplifier for CVD-diamond particle detectors},
journal = {Journal of Instrumentation}
}

@article{BORTFELDT2018317,
title = {PICOSEC: Charged particle timing at sub-25 picosecond precision with a Micromegas based detector},
journal = {Nuclear Instruments and Methods in Physics Research Section A: Accelerators, Spectrometers, Detectors and Associated Equipment},
volume = {903},
pages = {317-325},
year = {2018},
issn = {0168-9002},
doi = {https://doi.org/10.1016/j.nima.2018.04.033},
url = {https://www.sciencedirect.com/science/article/pii/S0168900218305369},
author = {J. Bortfeldt and F. Brunbauer and C. David and D. Desforge and G. Fanourakis and J. Franchi and M. Gallinaro and I. Giomataris and D. González-Díaz and T. Gustavsson and C. Guyot and F.J. Iguaz and M. Kebbiri and P. Legou and J. Liu and M. Lupberger and O. Maillard and I. Manthos and H. Müller and V. Niaouris and E. Oliveri and T. Papaevangelou and K. Paraschou and M. Pomorski and B. Qi and F. Resnati and L. Ropelewski and D. Sampsonidis and T. Schneider and P. Schwemling and L. Sohl and M. van Stenis and P. Thuiner and Y. Tsipolitis and S.E. Tzamarias and R. Veenhof and X. Wang and S. White and Z. Zhang and Y. Zhou}
}

@misc{utrobicic2024singlechannelpicosecmicromegas,
      title={Single channel PICOSEC Micromegas detector with improved time resolution}, 
      author={A. Utrobicic and R. Aleksan and Y. Angelis and J. Bortfeldt and F. Brunbauer and M. Brunoldi and E. Chatzianagnostou and J. Datta and K. Dehmelt and G. Fanourakis and D. Fiorina and K. J. Floethner and M. Gallinaro and F. Garcia and I. Giomataris and K. Gnanvo and F. J. Iguaz and D. Janssens and A. Kallitsopoulou and M. Kovacic and B. Kross and P. Legou and M. Lisowska and J. Liu and M. Lupberger and I. Maniatis and J. McKisson and Y. Meng and H. Muller and E. Oliveri and G. Orlandini and A. Pandey and T. Papaevangelou and M. Pomorski and L. Ropelewski and D. Sampsonidis and L. Scharenberg and T. Schneider and L. Sohl and M. van Stenis and Y. Tsipolitis and S. E. Tzamarias and I. Vai and R. Veenhof and P. Vitulo and X. Wang and S. White and W. Xi and Z. Zhang and Y. Zhou},
      year={2024},
      eprint={2406.05657},
      archivePrefix={arXiv},
      primaryClass={physics.ins-det},
      url={https://arxiv.org/abs/2406.05657}, 
}

@article{Utrobicic_2023,
   title={A large area 100-channel PICOSEC Micromegas detector with time resolution at the 20 ps level},
   volume={18},
   ISSN={1748-0221},
   url={http://dx.doi.org/10.1088/1748-0221/18/07/C07012},
   DOI={10.1088/1748-0221/18/07/c07012},
   number={07},
   journal={Journal of Instrumentation},
   publisher={IOP Publishing},
   author={Utrobicic, A. and Angelis, Y. and Bortfeldt, J. and Brunbauer, F. and Chatzianagnostou, E. and Dehmelt, K. and Fanourakis, G. and Floethner, K.J. and Gallinaro, M. and Garcia, F. and Garg, P. and Giomataris, I. and Gnanvo, K. and Gustavsson, T. and Iguaz, F.J. and Janssens, D. and Kallitsopoulou, A. and Kovacic, M. and Legou, P. and Lisowska, M. and Liu, J. and Lupberger, M. and Maniatis, I. and Meng, Y. and Muller, H. and Oliveri, E. and Orlandini, G. and Papaevangelou, T. and Pomorski, M. and Ropelewski, L. and Sampsonidis, D. and Scharenberg, L. and Schneider, T. and Sohl, L. and van Stenis, M. and Tsipolitis, Y. and Tzamarias, S.E. and Veenhof, R. and Wang, X. and White, S. and Zhang, Z. and Zhou, Y.},
   year={2023},
   month=jul, pages={C07012} }

\end{document}